# A universal shape function for rising jets

Cees J.M. van Rijn[1], Willem G.N. van Heugten[1] and Egbert Boeker[2]

[1] Microfluidics and Nanotechnology, ORC, Wageningen University Research, Stippeneng 4, 6708WE Wageningen, The Netherlands,  cees.vanrijn@wur.nl

[2] Department of Physics and Astronomy, VU University - Faculty of Sciences, De Boelelaan 1081, NL-1081 HV    Amsterdam, The Netherlands, e.boeker@vu.nl

Abstract:
A small drop that splashes into a deep liquid sometimes reappears as a small rising jet, for example when a water drop splashes into a pool or when coffee drips into a cup. Here we describe that the growing and rising jet continuously redistributes its fluid to maintain a universal shape originating from a surface tension based deceleration of the jet; the shape is universal in the sense that the shape of the rising jet is the same at all times; only the scaling depends on fluid parameters and deceleration. An inviscid equation of motion for the jet is proposed assuming a time dependent but uniform deceleration; the equation of motion is made dimensionless by using a generalised time dependent capillary length $\lambda_c$ and is solved numerically. As a solution a concave shape function is found that is fully determined by three measurable physical parameters: deceleration, mass density and surface tension; it is found that the surface tension based deceleration of the jet scales quadratic with the size of the jet base. Deceleration values derived from the jet shape are in good agreement with deceleration values calculated from the time plot of the height of the rising jet.

The scientific history of rising jets starts in the 19th century [1,2] when A.M. Worthington observed that small liquid jets are formed after impact of a small falling object in water. Numerous studies on this phenomenon have been carried out since [3-7], focusing mainly on different fluid responses when a small object or droplet hits the water surface; for example coalescence, crown splashing and central jet formation [3,8-12]. With respect to jet formation Gekle et al. [10] showed with numerical simulations that the jet acquires an upward momentum in a relatively small upward acceleration region located around the jet base due to implosion of a cavity created after drop impact. In further work [4] two more jet regions were distinguished: a long ballistic region, where fluid moves with a constant velocity and a jet tip region where surface tension and breakup effects dominate. The evolution and shape of such a jet was described by considering a radial dependent initial velocity profile (line of sinks) at the base of the jet [4,10]. In the present paper we find that the growing and rising jet continuously redistributes its fluid to maintain a universal shape originating from a surface tension based deceleration of the jet. We do not need any assumption about the initial fluid velocity distribution.

The physical process is illustrated in Fig.1a, where snapshots taken from a coffee jet are shown at times between 3 ms and 20 ms after the outburst of the jet. The ballistic region is the region below the jet tip region where droplet formation is initiated. The varying height $H(t)$ of the tip of the jet is given in Fig.1b and the deceleration $a(t)$, being the second time derivative of the height plot ($a(t) \equiv -\partial^2 H(t)/\partial t^2$), is depicted with black dots in Fig.1c. The jet deceleration $a(t)$ drops quickly from 100-200 m/s$^2$ to a value of about 20 m/s$^2$ after 15-20 ms, a value well above the gravitational constant ($g$=9.81m/s$^2$). Besides gravity the jet will experience a surface tension based deceleration pulling the jet back to the liquid surface; the total deceleration $a(t)$ can then be considered as a sum of two parts $a(t)=d(t)+g$ with $d(t)$ the surface tension based deceleration. We will show that $d(t)$ is a shape parameter determining both shape and absolute size of the jet. The blue diamonds in Fig. 1c display $a=d+g$ as a function of time. After deriving the relevant equations we will discuss in detail how the blue diamonds are found from the experimental data. The agreement between black dots and blue diamonds strongly supports our simple model.

The main points of the paper are: (1) a new physical equation of motion for a rising jet derived from Newton's laws, (2) a universal function for the concave shape of a rising jet deduced from this equation, using mass density and surface tension as physical parameters, (3) deceleration $a(t)=d(t)+g$ as the only time dependent parameter determining the absolute size and shape of the jet, (4) verification that the universal function describes the shape of rising jets and (5) check that deceleration values $a(t)$ from the tip heights and from the time dependent shape parameter $d(t)$ match with each other both for small and large jets.

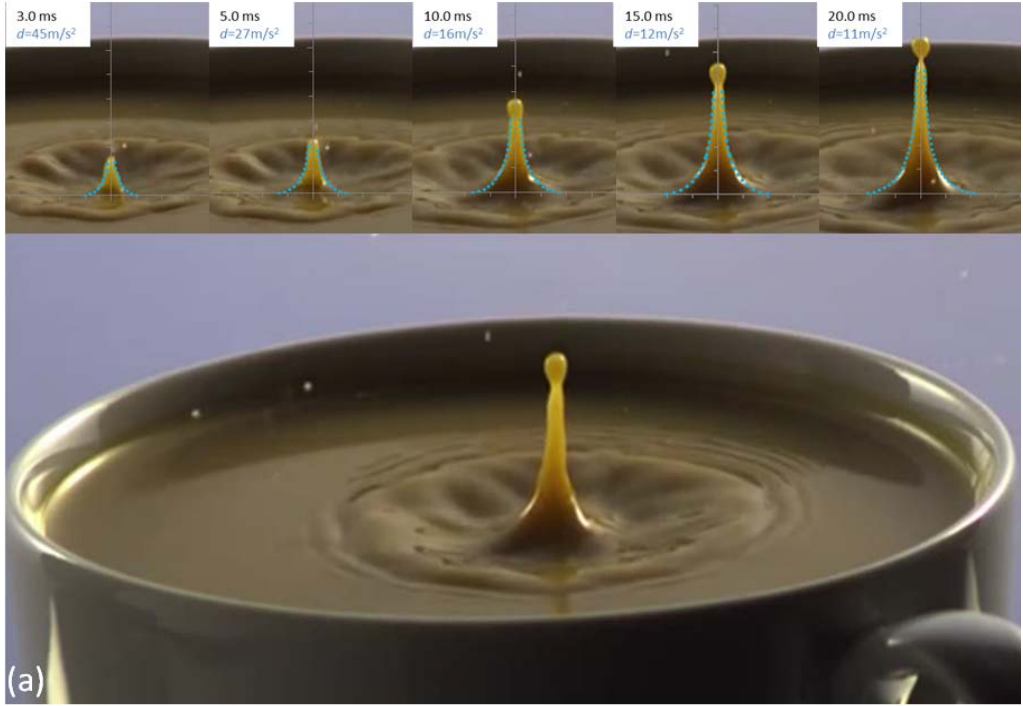

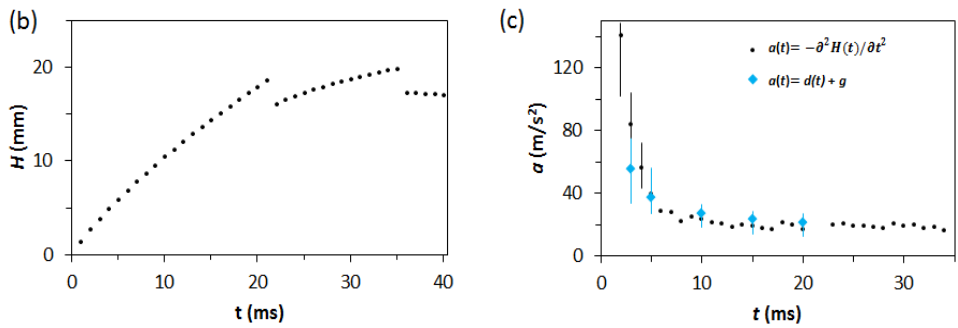

FIG 1. Snapshots of a coffee jet. (a), Snapshots at 3.0, 5.0, 10.0, 15.0 and 20.0 ms after initial outburst of the jet. Blue dotted lines correspond to the universal shape function (Fig.3a) with $\gamma$=50 mN/m, $\rho$=1000 kg/m$^3$. Scale bar is 2.5 mm. (b), Time plot of the jet height $H(t)$ measured from base to top of the jet. At 21.0 and 36.0 ms a droplets is emitted from the tip resulting in a discontinuity in $H(t)$. (c), Deceleration $a(t)$ of the jet obtained from the second time derivative ($-\partial^2 H(t)/\partial t^2$) of the time plot of $H$ (black dots), and obtained by fitting the shape function to the snapshots yielding for each jet a specific value for $d(t)$ in m/s$^2$; then $a(t)=d(t)+g$ (blue diamonds with error bars).

Note in Fig.1a that for all snapshots between 3 and 20 ms the same universal shape function (cf. Fig.3a) is depicted. Only the absolute size of the contour changes when the jet rises. The rising jet grows at its base due to arrival of new fluid emanating from the small upward acceleration region [4]. In the jet itself the surface tension determines the universal shape of the jet (cf. eq.(4)). This implies that at any moment some fluid needs to be redistributed both axially and radially to maintain that universal jet shape. Clearly this redistribution process will be delayed or hampered at increased viscosity values of the fluid as will be discussed below Fig.3a.

In order to derive the equations of motion a mathematical tool is used. The jet is subdivided into a sum of hollow cylinders, each cylinder experiencing a downward gravitational force and a downward surface tension force originating from the curved surface $\Delta A$ directly above that cylinder. Each hollow cylinder has an infinitesimal small thickness $\Delta R$, of which one, with height $h$ is shown in Fig. 2.

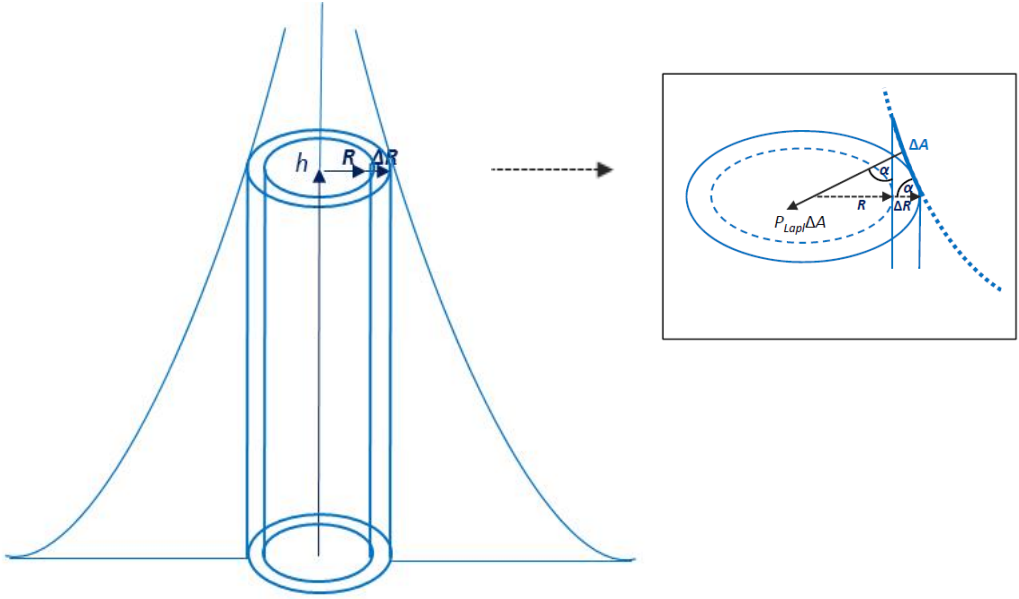

FIG 2. Rising jet subdivided in hollow cylinders. One of the cylinders is depicted. It has height $h$, radius $R$ and thickness $\Delta R$. Inset shows that at height $h$ the jet surface with a curved surface area $\Delta A$ exerts an inward surface tension based force $\Delta A 2\gamma/R_C$.

At height $h$ the outer radius of the hollow cylinder matches with the radius $R(h)$ of the jet. Each hollow cylinder experiences a gravitational force with value $m_{cyl}g$, with $m_{cyl}$ the mass of one cylinder with height $h$. The cylinder contacting the jet surface also experiences the surface tension at height $h$, as drawn in Fig. 2 (inset). The resulting acceleration will be a sum of two downward terms, gravity and surface tension, and is given by:

$$F = m_{cyl}\, a \equiv m_{cyl}\,(g+d) \tag{1}$$

which defines $d(t)$. In order to estimate the resulting force, consider a small circular curved surface element $\Delta A$ (inset Fig.2) with a mean radius of curvature $R_c$, which will be discussed briefly following eq.(5). Surface tension induces a Laplace pressure difference with value $P_{Lapl}=2\gamma/R_c(h)$ between both sides of the curved surface. The curved surface with area $\Delta A$ will exert an inward directed force with value $P_{Lapl}\Delta A$. If one projects this force on the (horizontal) top plane of the hollow cylinder with surface area $2\pi R\Delta R$ one finds that the downward force acting on the cylinder becomes $P_{Lapl}\Delta A\cos\alpha$. Therefore:

$$F = m_{cyl} a = m_{cyl} g + P_{Lapl} \Delta A \cos\alpha. \tag{2}$$

Using $m_{cyl} = \rho\, 2\pi R \Delta R\, h$ and $\Delta A \cos\alpha = 2\pi R \Delta R$ one gets

$$\rho a h = \rho g h + 2\gamma/R_c(h) \tag{3}$$

At any time $t$ a specific time dependent value for the total deceleration $a(t)$ of the jet holds. Equation of motion (3) can be simplified by using $d(t)=a(t)-g$. It will be shown that the shape of the jet is determined by $d(t)$ only. Therefore we call $d(t)$ the shape parameter, a parameter that quantitatively determines the absolute contour of the jet. The time dependent equation of motion for a rising liquid jet thus reads:

$$2\gamma/R_c(h) = \rho d(t) h \tag{4}$$

In deriving this equation it was assumed that viscosity between the hollow cylinders can be neglected and that all hollow cylinders experience the same deceleration $a$. No assumption about velocities inside the jet were made.

Equation (4) can be made dimensionless by a generalized time dependent capillary length $\lambda_c(t) \equiv \sqrt{(\gamma/\rho d(t))}$ if one defines $R^*_c \equiv R_c/\lambda_c$ and $h^* \equiv h/\lambda_c$. The dimensionless mean curvature $1/R^*_c(h^*)$ is a function of $h^*$ and determines the shape of the dimensionless radial coordinate of the liquid jet $R^*(h^*)$ along the dimensionless $h^* \equiv h/\lambda_c$ -axis. Possible functional shapes of $R^*(h^*)$ are determined by the following non-linear partial differential equation [13-15], which follows from equation (4):

$$h^* = \frac{2}{R^*_c(h^*)} \equiv \frac{1}{R^*_1(h^*)} + \frac{1}{R^*_2(h^*)} \equiv \frac{1}{R^*(h^*)\sqrt{1+[R^{*'}(h^*)]^2}} - \frac{R^{*''}(h^*)}{\left(1+[R^{*'}(h^*)]^2\right)^{3/2}} \tag{5}$$

Note that for a surface the mean curvature, which determines the Laplace pressure, should be written as a sum of two curvature terms $1/R^*_1(h^*)$ and $1/R^*_2(h^*)$ and that one of the terms is negative because of the concave shape of the jet.

For large values of height $h^*$ both $R^{*'}$ and $R^{*''}$ tend to zero; therefore the asymptotic solution becomes $R^*(h^*)=1/h^*$. This approximate solution can be used as starting point to solve eq. (5) numerically; we usually start at $h^* = 10$ with the approximate values as initial conditions. The dimensionless result is shown in Fig.3a, obtained with Mathematica$^{TM}$.

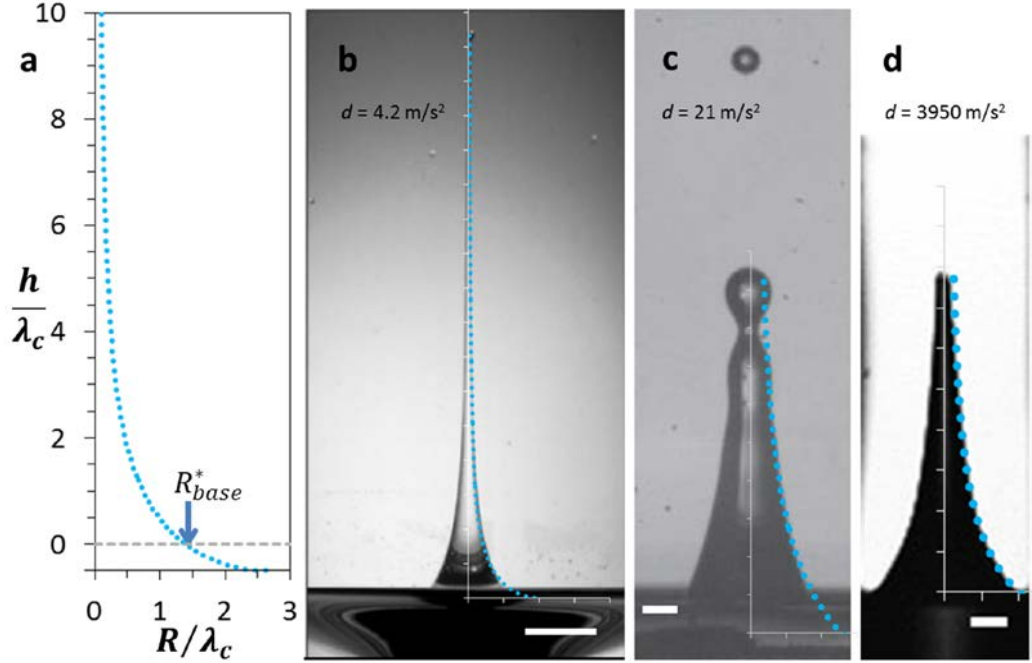

FIG 3. Concave jet shapes. (a), Dimensionless plot (blue dotted line) of the concave jet shape according to the numerical solution $R^*(h^*)$ of equation (5), Height $h^*$ and radius $R^*$ are given in dimensionless units $h/\lambda_c$ and $R/\lambda_c$ with $\lambda_c$ the capillary length, $R^*_{base}$ ≡$R^*(h^*$=0)=1.39  (b), Fitted dimensionfull jet shape according to the numerical solution $R(h)= \lambda_c\, R^*(h/\lambda_c)$  (blue dotted line) for a glycerol jet with viscosity 19.4 mPa.s from ref.[16]. Scale bar is 1 cm. (c), Idem as (b) for a pure water jet from ref.[17] after impact of a drop. Scale bar is 1 mm. (d), Idem as (b) for a high velocity micro jet from ref.[18]. Scale bar is 100 µm.

The fully concave solution of Eq. (5) is plotted in Fig.3a. In order to make this dimensionless solution dimensionfull one has to multiply both the horizontal and the vertical coordinates, respectively $h^*$ and $R^*$, with the same factor $\lambda_c$. This explains why the shapes of all jets of all liquids are the same. In the course of time the shape remains the same while the size increases.

In Table S1 dimensionless numerical values for the shape function $R^*$ and $h^*$ are listed.  The dimensionless value for the radius at the base is $R^*_{base} \equiv R^*(h^*$=0) =1.39 and at this point the mean curvature $1/R^*_c(0)$ is zero. Concave jet shapes are often described in the literature [4,10,16,17,18] without deriving the shape from the deceleration of the jet. In Fig.3b-d we show three of these jet shapes which resemble the shapes of Fig 1a. In all cases the values of the shape parameter $d$ follow from the fitting procedure.

The first step in the procedure is to use the fact that at the base ($h$=0) of the jet $R_{base}$≡1.39 $\lambda_c$. This gives a first estimate of the capillary length $\lambda_c$. Next enlarge the experimental photographs until individual pixels become visible; find the conversion factor between pixels and millimeters. Put down, by hand, the coordinates of the pixels of the interface between fluid and air.

Next use table S1 for the dimensionless graph and the conversion factor to find the dimensionfull graph for the first estimate of $\lambda_c$. Next try several values of the capillary length $\lambda_c$ around the first estimate to find the closest value that matches with the pixels. For that value a relative fit error for the radius $R$ is obtained by comparing the difference of the radius $R$ according to the experimental contour of the jet and of the theoretically obtained shape function using the formula $\Delta R/R = \sqrt{\sum_{i=1}^{i=n}(R_{i\,exp} - R_{i\,theory})^2 /(n \cdot R_{i\,theory}^2)}$. Typically the averaged error in $\Delta R/R$ is found to be 5-12% in Fig.1a, 10-15% in Fig. 3b,c,d, and 7-10% in Fig.4a. The relative fit error $\Delta d/d$ is estimated as about 1.5 times the value found for $\Delta R/R$. In this way $d$ and error bars $\Delta d/d$ as depicted in Fig 1c and 4c were found. We submit that the agreement between the deceleration found from fitting the shape and directly from the deceleration of the tip strongly supports the validity of our model.

In Fig.3b a jet is depicted, that is formed by the collapse of a singularity of a glycerol/water mixture (cf. Fig. 1 of ref.[16]). A tall standing wave is generated in a tank producing a deep depression/cavity, which subsequently collapses. We find a jet shape parameter $d=4.2\pm0.3$ m/s$^2$. Unfortunately no height plot $H(t)$ in time is available to calculate $a(t)= -\partial^2 H(t)/\partial t^2$) and to derive a second, and independent estimate for $d(t)= a(t)$-$g$. Note however that the shape function fits very well up to the tip of the jet ($\Delta R/R \leq 5\%$). The glycerol/water mixture used has a viscosity of 19.4 mPa.s; the model may therefore also apply to fluids more viscous than water. The internal redistribution of fluid within the jet to maintain its universal shape will however be hampered when the viscous time $\tau_{visc} \sim \eta l/\gamma$ to redistribute fluid over a typical length $l$ becomes too large. We observed that up to a viscosity of 50 mPa.s glycerol jets are still being formed and shaped according to Fig.3a after drop impact. The viscous time scale $\tau_{visc}$ for a typical radial or axial length $l$ of a few millimetres increases then to a few milliseconds, which is still small with respect to the time for the jet to rise to its apex (typical 2 cm in 50 ms).

In Fig.3c a low velocity jet (obtained after drop impact) is depicted, taken from Fig. 2e of ref.[17]. By fitting the shape of the jet by the blue dotted line one obtains the numerical solution for $R(h)$ with a shape parameter $d=21\pm5$ m/s$^2$. From the tip velocity plot as obtained from Fig. 2) of ref.[17] a corresponding jet tip deceleration can be derived with a value of $a = 38\pm5$ m/s$^2$. This leads to $d=a-g=28 \pm5$ m/s$^2$; this match in $d$ values within error margins again supports the model.

In Fig. 3d a high velocity water jet from Fig.3a of ref. [18] is depicted and fitted. Here a high energy laser pulse is used to create a vapour bubble, that subsequently generates a high velocity microjet (>10 m/s) with corresponding high Weber number (We>1000). We find here a shape parameter $d$ of about 3,950 m/s$^2$. This corresponds to $R_{base} \approx 200$ µm, which agrees with Fig.3d. A time plot of the velocity of the tip is presented in Fig. 3 of ref. [19] from which a deceleration can be derived with a value of about 5,000 m/s$^2$. This suggests that within a margin of 20% the model may also be applicable to high Weber number microjets.

It may be worthwhile to note that both Figs.3b and 3d, in contrast to Figs.1a and 3c depict rising jets without a visible jet tip region (where surface tension and breakup effects dominate with Rayleigh-Plateau instabilities). In a recent study [7] it is concluded that jet tip breakup is suppressed for Ohnesorge numbers larger than 0.091, regardless of the Weber number. A jet without a clear jet tip region that completely resembles the concave shape function, such as depicted in Figs.3b and 3d, will possibly fulfill these conditions [7].

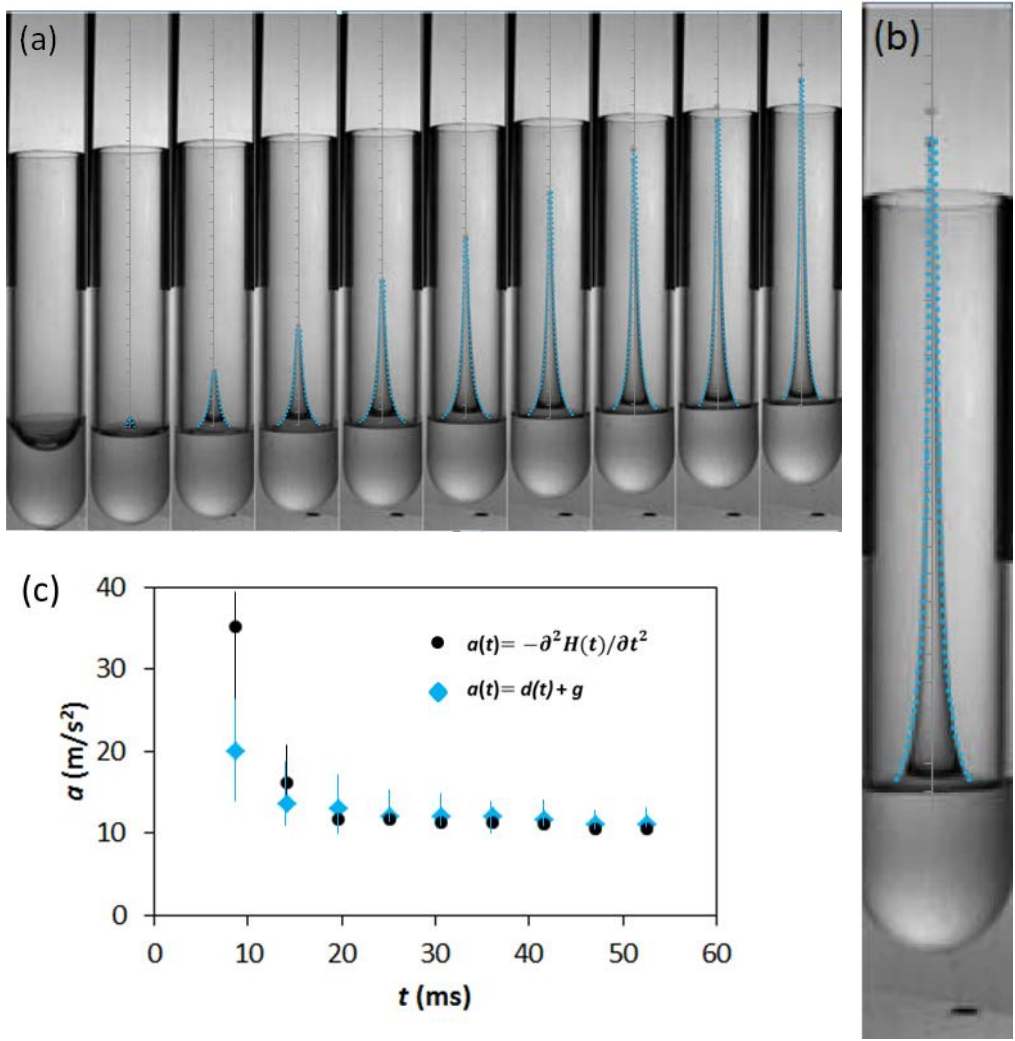

FIG 4. A glass tube is falling and at impact on a rigid floor a concentrated jet is formed, ref. [20]. (a), The interval separating each snapshot is 5.5ms. Blue dotted lines correspond to the universal shape function (Fig.3a) with $\gamma$=72.8 mN/m, $\rho$=1000 kg/m$^3$. Tube diameter is 3 cm. (b), Enlargement of the last snapshot of Fig.4a. (c), Deceleration $a(t)$ of the jet obtained from the second time derivative of the time plot of $H$ (black dots), and obtained by fitting the universal shape function to the snapshots yielding for each jet a specific value for $d(t)$ in m/s$^2$ according to $a(t)=d(t)+g$ (blue diamonds).

In the examples discussed so far the dimensions of the bath forming the jets were large with respect to the jet.

It is interesting to investigate whether the model may apply to other experimental setups. Therefore we studied in Fig.4 a series of photos made when a glass tube filled with water falls on a rigid floor [20]. When the tube hits the floor the hollow meniscus will flatten and the suddenly downward moving fluid at the glass edge will generate a fast rising jet in the centre of the tube with an initial velocity of ≈8m/s. The tube itself will recoil due to a partial elastic collision with the floor and has a much slower initial velocity(≈1.5m/s).

In Fig.4a 9 snapshots are depicted of the rising jets together with a blue dotted fit to the universal shape function (cf. Fig.3a). The averaged fit error $\Delta R/R$ is 7-10% for the last 8 snapshots. Fig.4b shows a magnification of the last snapshot at 52.7 ms after the impact on the floor; the value for the derived deceleration $d$ is 1.0±0.15 m/s$^2$. Fig.4c depicts with black dots the second time derivative of the height plot ($a(t) = -\partial^2 H(t)/\partial t^2$). Note that the jet deceleration $a(t)$ drops quickly from about 35 m/s$^2$ to 10-12 m/s$^2$ in less than 15-20 ms. The blue diamonds in Fig. 4c display $a=d+g$ as a function of time. The agreement between black dots and blue diamonds again strongly supports our simple model even in this experimental setup.

The fact, that the dimensionless shape of the jet, depicted in Fig. 3a, is independent of density and surface tension of the liquid and of the time of the measurement, leads to an interesting scaling phenomenon; from $R_{base}=1.39\lambda_c(t)$ and $\lambda_c(t) \equiv \sqrt{(\gamma/\rho d(t))}$ it follows that the deceleration $d(t)$ at any moment scales with the square of the radius of the jet base. This means that a smaller deceleration always corresponds to a larger base. Further the volume and mass of the jet scale cubic with the dimensionfull generalized capillary length $\lambda_c$ and thus with $R_{base}$. This means that a larger base always corresponds with a larger mass. These scaling laws apply independent of the liquid and hold for any time.

Finally we note that for a rising jet gravity and surface tension operate in the same direction, downwards. For a pending liquid thread, or honey slowly dripping from a spoon, the surface tension operates up and gravity down, while the acceleration is zero. Consequently equation (4) will apply with $a=0$ and $d=g$. The corresponding radius at the jet base would be $R_{base} \equiv 1.39 \lambda_c = 1.39\sqrt{(\gamma/\rho g)}$.

We conclude that the simple model, using only gravity and surface tension as physical forces works quite well for a variety of jets. The model may also be relevant for other applications wherein inertial acceleration and interfacial effects occur simultaneously, such as tail evolution of inkjet drops [22,23], solid nanojets formed with pulsed laser deposition [24], and free-surface singular jet phenomena [21].

**Supplemental Materials:**

TABLE S1.

Numerically obtained dimensionless values for h, R, R", $2/R_c$, $1/R_1$ and $1/R_2$ for concave solution Fig.3a of equation (4). The dimensionless volume of the jet is defined as $V_{jet}(h) = \int_h^\infty \pi R^2(x) dx$.

| h | R | R" | $2/R_c$ | $1/R_1$ | $1/R_2$ | $V_{jet}$ |
|---|---|---|---|---|---|---|
| ~-0.498 | ~2.7 | | | | | ~10.982 |
| -0.497 | 2.620 | >13000 | ~-0.497 | 0.013 | -0.484 | 10.98 |
| -0.49 | 2.486 | 806.34 | -0.49 | 0.035 | -0.525 | 10.86 |
| -0.48 | 2.397 | 219.15 | -0.48 | 0.056 | -0.533 | 10.67 |
| -0.47 | 2.332 | 111.04 | -0.47 | 0.073 | -0.541 | 10.49 |
| -0.46 | 2.278 | 69.58 | -0.46 | 0.087 | -0.546 | 10.33 |
| -0.45 | 2.232 | 49.18 | -0.45 | 0.100 | -0.553 | 10.17 |
| -0.44 | 2.191 | 36.61 | -0.44 | 0.113 | -0.551 | 10.01 |
| -0.43 | 2.153 | 28.95 | -0.43 | 0.124 | -0.555 | 9.86 |
| -0.42 | 2.119 | 23.56 | -0.42 | 0.135 | -0.556 | 9.72 |
| -0.41 | 2.086 | 19.61 | -0.41 | 0.146 | -0.555 | 9.58 |
| -0.4 | 2.056 | 16.74 | -0.40 | 0.156 | -0.557 | 9.45 |
| -0.3 | 1.819 | 5.90 | -0.30 | 0.249 | -0.549 | 8.27 |
| -0.2 | 1.646 | 3.27 | -0.20 | 0.332 | -0.534 | 7.33 |
| -0.1 | 1.507 | 2.15 | -0.10 | 0.411 | -0.510 | 6.55 |
| 0 | 1.390 | 1.56 | 0.00 | 0.488 | -0.487 | 5.89 |
| 0.1 | 1.289 | 1.19 | 0.11 | 0.563 | -0.452 | 5.33 |
| 0.2 | 1.199 | 0.99 | 0.19 | 0.638 | -0.445 | 4.84 |
| 0.3 | 1.120 | 0.83 | 0.29 | 0.714 | -0.424 | 4.42 |
| 0.4 | 1.049 | 0.69 | 0.40 | 0.791 | -0.394 | 4.05 |
| 0.5 | 0.984 | 0.58 | 0.51 | 0.867 | -0.362 | 3.72 |
| 0.6 | 0.926 | 0.50 | 0.61 | 0.943 | -0.334 | 3.44 |
| 0.7 | 0.873 | 0.45 | 0.70 | 1.021 | -0.318 | 3.18 |
| 0.8 | 0.824 | 0.43 | 0.78 | 1.100 | -0.316 | 2.96 |
| 0.9 | 0.779 | 0.36 | 0.90 | 1.180 | -0.279 | 2.75 |
| 1 | 0.738 | 0.32 | 1.00 | 1.260 | -0.262 | 2.57 |
| 1.1 | 0.700 | 0.29 | 1.10 | 1.342 | -0.244 | 2.41 |
| 1.2 | 0.665 | 0.26 | 1.20 | 1.425 | -0.226 | 2.26 |
| 1.3 | 0.633 | 0.24 | 1.30 | 1.509 | -0.208 | 2.13 |
| 1.4 | 0.603 | 0.22 | 1.40 | 1.593 | -0.192 | 2.01 |
| 1.5 | 0.575 | 0.20 | 1.50 | 1.679 | -0.178 | 1.90 |
| 1.6 | 0.550 | 0.18 | 1.60 | 1.766 | -0.166 | 1.80 |
| 1.7 | 0.526 | 0.17 | 1.70 | 1.853 | -0.154 | 1.71 |
| 1.8 | 0.504 | 0.15 | 1.80 | 1.942 | -0.142 | 1.63 |
| 1.9 | 0.483 | 0.14 | 1.90 | 2.031 | -0.131 | 1.55 |
| 2 | 0.463 | 0.13 | 2.00 | 2.121 | -0.119 | 1.48 |
| 2.1 | 0.445 | 0.12 | 2.10 | 2.212 | -0.112 | 1.42 |
| 2.2 | 0.428 | 0.11 | 2.20 | 2.303 | -0.104 | 1.36 |
| 2.3 | 0.413 | 0.10 | 2.30 | 2.395 | -0.096 | 1.30 |
| 2.4 | 0.398 | 0.090 | 2.40 | 2.488 | -0.088 | 1.25 |
| 2.5 | 0.384 | 0.083 | 2.50 | 2.582 | -0.080 | 1.20 |
| 2.6 | 0.371 | 0.078 | 2.60 | 2.676 | -0.076 | 1.16 |
| 2.7 | 0.358 | 0.073 | 2.70 | 2.770 | -0.071 | 1.12 |
| 2.8 | 0.347 | 0.066 | 2.80 | 2.865 | -0.065 | 1.08 |
| 2.9 | 0.336 | 0.059 | 2.90 | 2.960 | -0.058 | 1.04 |
| 3 | 0.326 | 0.057 | 3.00 | 3.056 | -0.056 | 1.01 |
| 4 | 0.248 | 0.027 | 4.00 | 4.027 | -0.026 | 0.75 |
| 5 | 0.199 | 0.015 | 5.00 | 5.015 | -0.015 | 0.60 |
| 6 | 0.166 | 0.011 | 6.00 | 6.010 | -0.011 | 0.49 |
| 7 | 0.143 | 0.007 | 7.00 | 7.007 | -0.007 | 0.42 |
| 8 | 0.125 | 0.003 | 8.00 | 8.003 | -0.003 | 0.36 |
| 9 | 0.111 | 0.002 | 9.00 | 9.002 | -0.002 | 0.32 |
| 10 | 0.100 | 0.001 | 10.00 | 10.001 | -0.001 | 0.28 |